\documentclass[10pt,conference]{IEEEtran}
\usepackage{minibox}
\usepackage{amsmath}
\usepackage{amssymb}
\usepackage{listings}
\usepackage{marvosym}
\usepackage{color}
\usepackage{graphicx}
\usepackage{subcaption}
\usepackage{pgf}
\usepackage{tikz}
\usetikzlibrary{arrows,automata,positioning,trees,shapes}
\usepackage{fancyvrb}
\usepackage{setspace}
\usepackage[linesnumbered,lined,algoruled]{algorithm2e}
\usepackage{url}
\usepackage{hyperref}
\usepackage[color=yellow!60,textsize=footnotesize]{todonotes}

\newenvironment{CVerbatim}
 {\singlespacing\center\BVerbatim}
 {\endBVerbatim\endcenter}

\ifCLASSINFOpdf
\else
\fi

\renewcommand{\baselinestretch}{0.95}

\begin{document}
%
\title{SMT-Based Refutation of Spurious Bug Reports in the Clang Static Analyzer}

\author{
\IEEEauthorblockN{
 Mikhail R. Gadelha\authorrefmark{1},
 Enrico Steffinlongo\authorrefmark{1},
 Lucas C. Cordeiro\authorrefmark{2},
 Bernd Fischer\authorrefmark{3}, and
 Denis A. Nicole\authorrefmark{1}
}
\IEEEauthorblockA{
 \authorrefmark{1}University of Southampton, UK.
 \authorrefmark{2}University of Manchester, UK.
 \authorrefmark{3}Stellenbosch University, South Africa.
}
}

%


%


\maketitle

\begin{abstract}
We describe and evaluate a bug refutation extension for the Clang Static
Analyzer (CSA) that addresses the limitations of the existing
built-in constraint solver.  In particular, we complement CSA's existing
heuristics that remove spurious bug reports. We encode the path constraints
produced by CSA as Satisfiability Modulo Theories (SMT) problems, use
SMT solvers to precisely check them for satisfiability, and remove bug reports
whose associated path constraints are unsatisfiable.
Our refutation extension refutes spurious bug reports in 8 out of 12
widely used open-source applications; on average, it refutes ca.\ 7\%
of all bug reports, and never refutes any true bug report.
It incurs only negligible performance overheads, and on average
adds 1.2\% to the runtime of the full Clang/LLVM toolchain.
A demonstration is available at {\tt https://www.youtube.com/watch?v=ylW5iRYNsGA}.
\end{abstract}



%
\IEEEpeerreviewmaketitle

\section{Introduction}


LLVM comprises a set of reusable components for program
compilation~\cite{Lattner:2004:LCF:977395.977673};
unlike other popular compilers, e.g., GCC and its monolithic
architecture~\cite{Novillo06gcc-an}. Clang~\cite{CLANG} is an LLVM
component that implements a
frontend for C, C++, ObjectiveC and their various extensions. Clang and LLVM
are used as the main compiler technology in several closed- and
open-source ecosystems, including
MacOS and OpenBSD~\cite{clang-openbsd}.

The Clang Static Analyzer (CSA)~\cite{7835996} is an open-source project built on top
of Clang that can perform context-sensitive interprocedural analysis for
programs written in the languages supported by Clang. CSA
symbolically executes the program, collects constraints, and reasons
about bug reachability using a built-in constraint solver.
It was designed to be fast, so that it can provide results for common mistakes
(e.g., division by zero or NULL pointer dereference) even in complex
programs. However, its speed comes at the expense of precision, and it cannot
handle some arithmetic (e.g., remainder) and bitwise operations. In
such cases, CSA can explore execution paths along which constraints do
not hold, which can lead to incorrect results being reported.
\begin{figure} [!ht]
\centering
  \begin{subfigure}{.45\textwidth}
\begin{lstlisting}[escapechar=|]
unsigned int func(unsigned int a) {
  unsigned int *z = 0;
  if ((a & 1) && ((a & 1) ^ 1)) |\label{clang:guard1}|
    return *z; |\label{clang:deref}|
  return 0;
}\end{lstlisting}
\end{subfigure}
\caption{A small C safe program. The dereference in line~\ref{clang:deref}
is unreachable because the guard in line~\ref{clang:guard1} is always false.}
\label{fig:clang-example-code}
\end{figure}

Consider the program in Fig.~\ref{fig:clang-example-code}. This
program is safe, i.e., the unsafe pointer dereference in line~\ref{clang:deref}
is unreachable because the guard in line~\ref{clang:guard1} is not satisfiable;
\texttt{a \& 1} holds if the last bit in \texttt{a} is one, and
\texttt{(a \& 1) \^{}  1} inverts the last bit in \texttt{a}. The analyzer,
however, produces the following (spurious) bug report when analyzing the
program:

\begin{CVerbatim}
main.c:4:12: warning: Dereference of null
 pointer (loaded from variable 'z')
    return *z;
           ^~
1 warning generated.
\end{CVerbatim}

The null pointer dereference reported here means that CSA claims to
nevertheless have found a path where the dereference of \texttt{z} is reachable.

Such spurious bug reports are in practice common; in our experience,
about 50\% of the reports in large systems are actually spurious.
Junker et~al.~\cite{Junker:2012:SFP:2428368.2428399} report similar numbers for
a similar static analysis technology.  Identifying
spurious bug reports and refactoring the code to suppress them puts a large
burden on developers and runs the risk of introducing actual bugs;
these issues negate the purpose of a lightweight, fast static analysis technology.

Here we present a solution to this conundrum. We first use the fast but
imprecise built-in solver to analyze the program and find potential bugs,
then use slower but precise SMT solvers to refute (or validate) them; a bug
is only reported if the SMT solver confirms that the bug is reachable.
We implemented
this approach inside Clang and evaluated it over twelve widely used C/C++
open-source projects of various size
using five different SMT solvers.
Our experiments show that our
refutation extension can remove false bug reports from 8 out of the 12 analyzed
projects; on average, it refuted 11 (or approximately 7\% of all) bug reports
per project, with a maximum of 51 reports refuted for XNU; it
never refuted any true bug report. Its performance overheads are negligible
and on average our extension adds only 1.2\% to the runtime of the full
Clang/LLVM toolchain.

\section{The Clang Static Analyzer}

CSA performs a context-sensitive interprocedural
data-flow analysis via graph reachability~\cite{Reps:1995:PID:199448.199462} and
relies on a set of checkers, which implement the logic for detecting specific
types of bugs~\cite{7835996,Horvath:2018:IEC:3183440.3195041}. Each path in a function
is explored, which includes taking separate branches and different loop
unrollings. Function calls on these paths are inlined whenever possible, so
their contexts and paths are visible from the caller's context.

Real-world programs, however, usually depend on external factors, such as user
inputs or results from library components, for which source code is not always
available~\cite{clang-static-guide}. These unknown values are represented by
algebraic symbols, and the built-in constraint solver in the static analyzer
reasons about reachability based on expressions containing these symbols.

CSA relies on a set of checkers that are
engineered to find specific types of bugs, ranging from undefined behaviour to
security property violations, e.g., incorrect usage of insecure
functions like \texttt{strcpy} and \texttt{strcat}~\cite{clang-static-checkers}.
The checkers subscribe to events (i.e., specific operations that occur
during symbolic execution) that are relevant to their bug targets;
for example, the nullability checker subscribes to pointer dereferences.
They then check the constraints in the current path and throw
warnings if they consider a bug to be reachable.


These checkers can report incorrect results since the symbolic analysis
is incomplete (i.e., they can miss some true bugs) and unsound (i.e., they can
generate spurious bug reports).  The sources of these incorrect results
are approximations in two components, namely the
control-flow analysis and the constraint solver.


The control-flow analysis evaluates function calls inside the same
translation unit (TU); if the symbolic execution engine finds a function call
implemented in another TU, then the call is skipped and pointers
and references passed as parameters are invalidated while the function return
value is assumed to be unknown. Cross translation unit support (CTU) is under
development~\cite{Horvath:2018:IEC:3183440.3195041}; it is not part of the
CSA main branch yet.

The built-in constraint solver (based on interval arithmetic) was built to
be fast rather than precise, and removes expressions from the reasoning
if they contain unsupported operations (e.g, remainders and bitwise operations) or
are too complex (e.g., contain more than one operator symbol).

The bug reports generated by the checkers are then post-processed before
they are reported to the user. In this final step, a number of heuristics
are applied to remove incorrect bug reports and to
beautify the reports. The reports are also deduplicated, so that
different paths that lead to the same bug only generate one report.
\begin{figure}[t]
  \centering
  \begin{tikzpicture}[>=stealth',shorten >=1pt,auto,node distance=2cm]
  \node[initial,state]   (s)                               {$s_0$};
  \node[state]           (s4)   [right of=s]               {$s_1$};
  \node[state]           (s1)   [right of=s4]              {$s_2$};
  \node[accepting,state] (s2)   [above of=s1]        {$s_3$};
  \node[state]           (s3)   [below of=s1]        {$s_4$};
  \node[accepting,state] (se1)  [right of=s3]              {$\epsilon$};

   \path[->] (s)  edge                       node {\texttt{a = *}} (s4)
             (s4) edge                       node {\texttt{z = 0}} (s1)
             (s1) edge [left]                node {\texttt{[(a \& 1) \&\& ((a \& 1) \^{} 1)]}} (s2)
             (s1) edge [left]                node {\texttt{[!((a \& 1) \&\& ((a \& 1) \^{} 1))]}} (s3)
             (s3) edge                       node {\texttt{*z}} (se1)
    ;
\end{tikzpicture}
\caption{Exploded graph of the program in Fig.~\ref{fig:clang-example-code},
which contains bitwise operations that are not properly handled by CSA's
original constraint solver, thus generating a spurious bug report.}
\vspace{-.5cm}
\label{fig:clang-example-graph}
\end{figure}
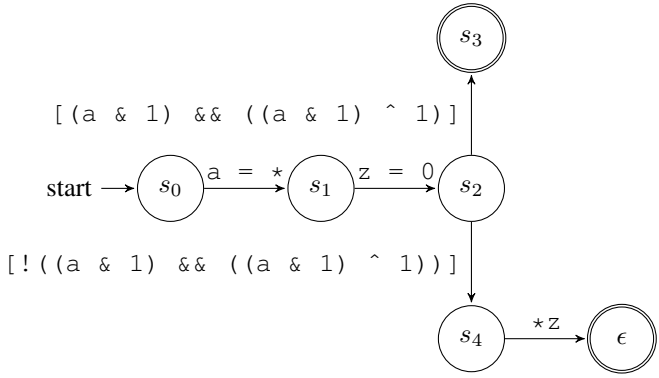

As an example of this process, consider the exploded
graph~\cite{Reps:1995:PID:199448.199462} in
Fig.~\ref{fig:clang-example-graph}; it represents the graph explored by the
analyzer when analyzing the program in Fig.~\ref{fig:clang-example-code}.
Here, a node $s$ is a pair $(V, C)$, where $V$ is a map $var \rightarrow
2^{128}$ that maps a value to every variable $var$ in the program, and $C$ is
a map $var \rightarrow 2^{128} \times 2^{128}$ that maps the interval
constraints to every variable $var$ required to reach that node. An edge is a
tuple $(s_i, \mathtt{OP}, s_{i+1})$, modeling the constraints on a transition
from $s_i$ to $s_{i+1}$, and $\mathtt{OP}$ is a operation performed in the
transition, either changing a constraint $c \in C$ or a read/write operation
over a $var \in V$. We also reserve two special symbols: an $\epsilon$ node is a
property violation and unknown values are shown as ``\texttt{*}'' symbols.

CSA's ``NullDereference'' checker static analyzer adds the transition to
$\epsilon$ representing a dereference of \texttt{z} when it is null.
Note that error node is added when the parameter \texttt{a} is read (even
if it was not explicitly initialized), because parameters are assumed to be
unknown rather than of uninitialized values, unless the control-flow analysis
infers otherwise.
The analyzer will explore all paths in the graph because it cannot infer that
the constraint that leads to node $s_4$ is always false.
%
\begin{figure}
  \centering
  \includegraphics[trim={4cm 2.5cm 1cm 2.5cm}, clip,
width=.55\textwidth]{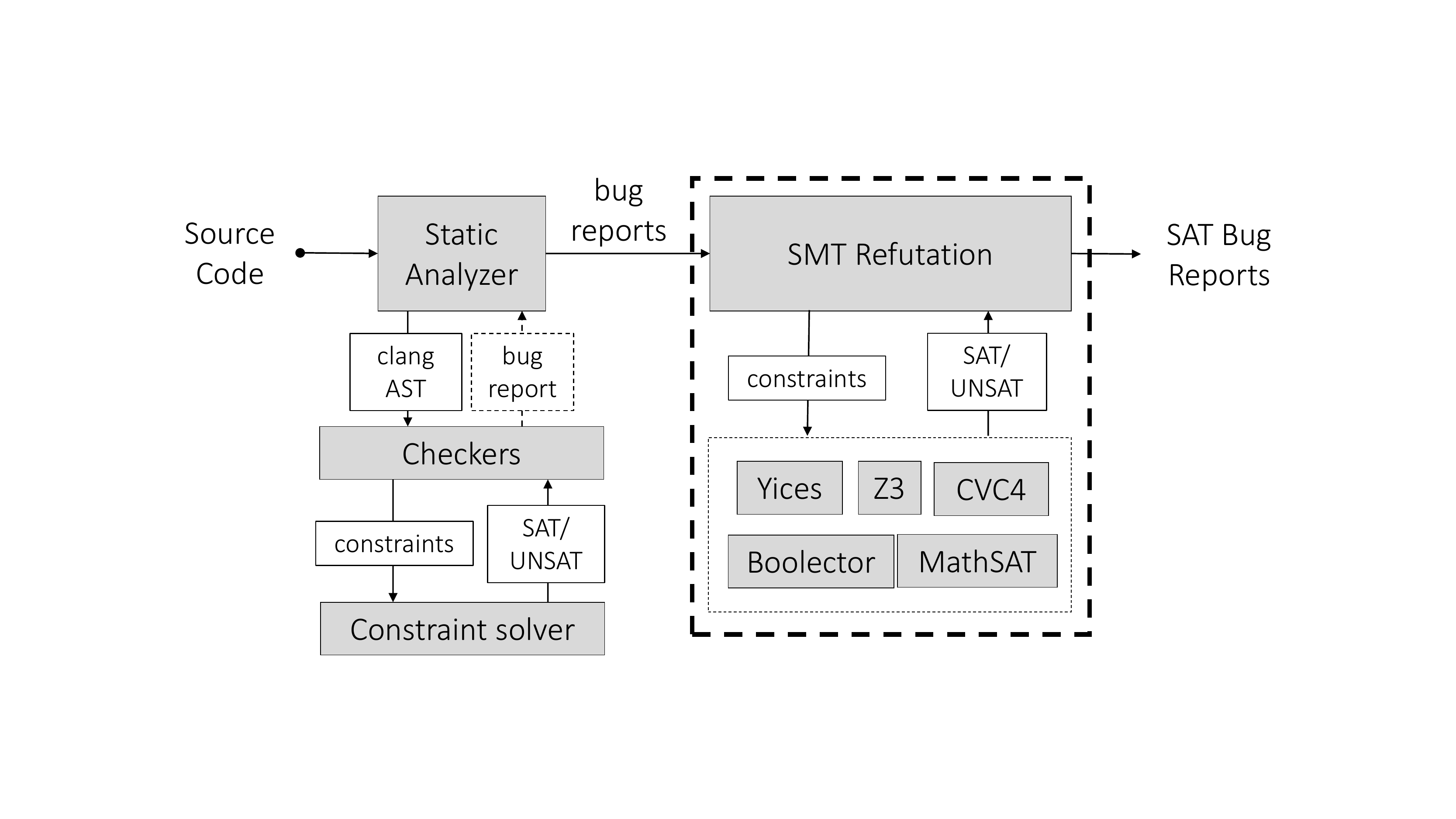}
\caption{The refutation extension in the Clang Static Analyzer.}
\label{fig:clang-arch}
\end{figure}

\section{Refuting False Bugs using SMT Solvers}
\label{sec:proposed-approach}

One approach to address the limitations of the built-in constraint solver is to
replace it by an SMT solver.  This approach has been implemented in
Clang but empirical evaluations show that this approach can be up to 20 times
slower.\footnote{\url{https://reviews.llvm.org/D28952}}


We developed an alternative solution: we use the more precise SMT
solvers to reason about bug reachability only
in the post processing step. CSA already has heuristics in place to remove incorrect
bug reports, so we extended those heuristics to precisely encode the constraints in
SMT and to check for satisfiability. Fig.~\ref{fig:clang-arch} illustrates the
architecture of our solution. After the static analyzer generates the
bug reports, the SMT-based refutation extension will encode the constraints as SMT formulas and
check them for satisfiability. CSA already supports constraint encoding
in SMT using Z3~\cite{Z08} but we also implemented support for
Boolector~\cite{Brummayer:2009:BES:1532891.1532912},
Yices~\cite{Yices}, MathSAT~\cite{Cimatti:2013:MSS:2450387.2450400}, and
CVC4~\cite{Barrett:2011:CVC:2032305.2032319}.

A bug report $\mathit{BR}$ is a straight line graph representing the path to a
property violation (i.e., an $\epsilon$-node).  Our refutation extension walks
backwards through all nodes $s_i$ in $\mathit{BR}$, collects their constraints,
and checks their conjunction for satisfiability. If the formula is
unsatisfiable, the bug is spurious.


Our constraints encoding algorithm is shown in
Algorithm~\ref{alg:encodeConstraint}. Assume a set of constraints $C$, an
SMT formula $\Phi$, and a method $encode(expr, \Phi)$, which encodes an
expression $expr$ in the SMT formula $\Phi$.
Algorithm~\ref{alg:encodeConstraint} contains two optimizations when encoding
constraints: duplicated symbol constraints are ignored
(line~\ref{clang:optimization1}) and if the constraint is a concrete value
(the lower bound is equal to the upper bound), the constraint is encoded as an
equality (line~\ref{clang:optimization2}).
Note that ignoring symbol constraints in line~\ref{clang:optimization1} is only
possible because the refutation extension runs from the last node in a bug
report (the property violation) to the initial node; any new symbol constraint
found when walking backwards will always be a subset of the symbol constraints
already encoded.

Fig.~\ref{fig:smt-refutation} shows the SMT formula of the bug found when
analyzing the program in Fig.~\ref{fig:clang-example-code}. The formula is
equivalent to the path $[s_0, s_1, s_2, s_4, \epsilon]$ in
Fig.~\ref{fig:clang-example-graph} and \texttt{\$0} is the value of the
variable \texttt{a}. Since the formula is unsatisfiable, CSA
will not produce a bug report for this path.
\begin{figure}[!h]
\centering
  \begin{subfigure}{.47\textwidth}
  \begin{lstlisting}[escapechar=^]
(declare-fun $0 () (_ BitVec 32))
(assert (= ((_ extract 0 0) $0) #b1))
(assert (= ((_ extract 0 0) $0) #b0))
  \end{lstlisting}
  \end{subfigure}
\caption{The SMT formula of the bug report from
Fig.~\ref{fig:clang-example-code}, using Z3. Note that the solver was able to
simplify the formula to two assertions: the first bit should be one and zero at
the same time. Since this is a contradiction, the formula is UNSAT.}
\label{fig:smt-refutation}
\end{figure}

\vspace{-.5cm}
\subsection{Running the Clang Static Analyzer}
\label{tool-instructions}

To run the clang static analyzer is enough to use the \texttt{scan-build}
tool shipped with clang. The tool runs the clang static analyzer during the
compilation of the project and it is simple to use: instead of running
\texttt{make}, simply run \texttt{scan-build make}. Scan-build offers several
options to customize the analysis, including enabling (and disabling) the
various checker in the clang static analyzer and enabling our refutation
extension. Once the build is done, a detailed report is generated for each
reported bug.

To analyze a project with our refutation extension enabled, run
\texttt{scan-build -analyzer-config `crosscheck-with-smt=true' make}. Detailed
instructions on how to run the clang static analyzer with the different solvers
are available in \url{https://github.com/mikhailramalho/clang}.

\section{Experimental Evaluation}
\label{clang:experimental-evaluation}

The experimental evaluation of our refutation extension consists of two parts.
We first present the research questions, projects evaluated and
the environment setup,
before we compare the analysis results of the CSA with and without the
bug refutation extension enabled; the two approaches are compared in terms of
number of refuted bugs and verification time.

\begin{algorithm}[!t]
\SetKw{Continue}{continue}
\KwIn{A set of constraints $C$ and an SMT formula $\Phi$}
\KwOut{The formula $\Phi$ with all constraints $c$ encoded in SMT}
\ForEach {$c \in C$}{
  \lIf{$c.var \in \Phi$}{\Continue} \label{clang:optimization1}

  \If{$c.interval.lower == c.interval.upper$}{ \label{clang:optimization2}
    $encode(c.var == c.interval.lower, \Phi)$
  }
  \Else{
  $encode(c.var \geq c.interval.lower \wedge \newline
  c.var \leq c.interval.upper, \Phi)$
  }
}
\caption{$encodeConstraint(cs, \Phi)$}
\label{alg:encodeConstraint}
\end{algorithm}

\subsection{Experimental Objectives and Setup}
\label{ref-benchmarks-description}

Our experimental evaluation aims to answer
two research questions:
\begin{enumerate}

\item[RQ1] \textbf{(soundness)} Is our approach sound and can the refuted bugs be confirmed?

\item[RQ2] \textbf{(performance)} Is our approach able to refute spurious bug reports in a reasonable amount of time?

\end{enumerate}

We evaluated the new bug refutation extension in twelve open-source C/C++
projects:
%
\textit{tmux (v2.7)}, a terminal multiplexer;
\textit{redis (v4.0.9)}, an in-memory database;
\textit{openSSL (v1.1.1-pre6)}, a software library for secure communication;
\textit{twin (v0.8.0)}, a windowing environment;
\textit{git (v2.17.0)}, a version control system;
\textit{postgreSQL (v10.4)}, an object-relational database management system;
\textit{SQLite3 (v3230100)}, a relational database management system;
\textit{curl (v7.61.0)}, command-line tool for transferring data;
\textit{libWebM (v1.0.0.27)}, a WebM container library;
\textit{memcached (v1.5.9)}, a general-purpose distributed memory caching system;
\textit{xerces-c++ (v3.2.1)}, a validating XML parser; and
\textit{XNU (v4570.41.2)}, the operating system kernel used in Apple products.

All experiments were conducted on a desktop computer with an Intel Core i7-2600
running at 3.40GHz and 24GB of RAM. We used Clang v7.0. A time limit of 15s
per bug report was set for the projects. All scripts required to analyze
all the projects are available in
\url{https://github.com/mikhailramalho/analyzer-projects}.

\subsection{Bug Refutation Comparison}
\label{ref-experimental-results}

\begin{table}[!h]
\begin{center}
\small
  \begin{tabular}{|l|c|c|c|c|}
    \hline
    Projects   & time (s) & time (s)   & reported bugs  & refuted \\
               & (no ref) & (with ref) & (no ref)       & bugs \\ \hline
    tmux       & 86.5     & 89.9       & 19             & 0  \\ \hline
    redis      & 347.8    & 338.3      & 93             & 1  \\ \hline
    openSSL    & 138.0    & 128.0      & 38             & 2  \\ \hline
    twin       & 225.6    & 216.7      & 63             & 1  \\ \hline
    git        & 488.7    & 405.9      & 70             & 11 \\ \hline
    postgreSQL & 1167.2   & 1112.4     & 196            & 6  \\ \hline
    SQLite3    & 1078.6   & 1058.4     & 83             & 15 \\ \hline
    curl       & 79.8     & 79.9       & 39             & 0  \\ \hline
    libWebM    & 43.9     & 44.2       & 6              & 0  \\ \hline
    memcached  & 96.0     & 96.2       & 25             & 0  \\ \hline
    xerces-c++ & 489.8    & 433.2      & 81             & 2  \\ \hline
    XNU        & 3441.7   & 3405.1     & 557            & 51 \\ \hline \hline
    Total      & 7683.7   & 7408.5     & 1270           & 89 \\ \hline
  \end{tabular}
\end{center}
\caption{Results of the analysis with and without refutation.}
\label{table:results}
\end{table}

Table~\ref{table:results} shows the results of CSA with and without bug refutation enabled. Here, \textit{time (s) (no
ref)} is the analysis time without refutation, \textit{time (s)
(ref)} is the analysis times with refutation enabled, averaged over
all supported solvers (Z3, Boolector, MathSAT, Yices and CVC4),
\textit{reported bugs (no ref)} is the number of bug reports produced without
refutation and \textit{refuted bugs} is the number of refuted bugs. All solvers
refuted the same bugs. There were bugs refuted in $8$ out of the $12$ analyzed
projects: redis, openssl, twin, git, postgresql, sqlite3, xerces and XNU.
On average, 11 bugs were refuted when analyzing these projects, with up to 51
bugs refuted in XNU.

In total, 89 bugs were refuted and an in-depth analysis of them show that all
of them were false positives, and thus affirm RQ1. Our technique, however, is not able to refute all
false bugs as the unsound interprocedural analysis is another source of false
positives in CSA and it was not addressed in this work.

The average time to analyze the projects with refuted bugs was 35.0 seconds
faster, a 6.25\% speed up, and thus affirm RQ2. This because the static analyzer generates html
reports for each bug, which involves intensive use of IO (e.g., the html
report produced for the program in Fig.~\ref{fig:clang-example-code} is
around 25kB), and by removing these false bugs, fewer reports are generated and
the analysis is slightly faster. Out of the four projects, where no bug was
refuted (tmux, curl, libWebM and memcached), the analysis was 1.0 second
slower on average: a 1.24\% slowdown.

\section{Related Work}

Static analysis of programs has seen a great improvement in the last years and
in many cases it has been applied for the analysis of big real world projects.
Frama-C~\cite{Kirchner2015} is an extensible analysis framework based on
abstract interpretation for the C language. It also has a large number of
plugins for checking different program properties.
Infer~\cite{10.1007/978-3-319-17524-9_1} is an open-source static code analysis
tool used for the verification of the Facebook code base and on many mobile apps
(e.g., WhatsApp) and is adopted by many companies (e.g., Mozilla,
Spotify). Cppcheck~\cite{CPPChecker} focuses on detecting dangerous coding
practices and undefined behaviours. It was used for detecting many
vulnerabilities including a stack overflow in X.org.

The use of SMT solvers as backends for static analysis tools is a well known
approach. It has been adopted in the ESBMC~\cite{esbmc2018} C/C++ bounded model
checker. It encodes the program as a SMT formula and depending on its
satisfiability it detects possible bugs. This encoding technique is proved to be
sound, however, it does not scale up to real world programs.

An approach similar to the one adopted in this paper has also been used in
Goanna~\cite{Junker:2012:SFP:2428368.2428399}: a C/C++ static analyzer able to
scale up to real world programs (e.g., OpenSSL, WireShark). After detecting all
bugs, false ones are eliminated by analyzing the feasibility of error paths
using SMT solvers. The biggest difference between Goanna and our approach is
on the first (imprecise) analysis: Goanna uses NuXmv and MathSAT5 to generate
the bug reports, while we use a custom built constraint solver. Similarly to
our approach, they use Z3 to encode and refute false bug reports, but we offer a
wider selection of SMT solvers to choose from.

\section{Conclusions}

%
%
%
%

Our SMT-based bug refutation extension in the clang static analyzer is a simple but
powerful extension; it is able to prevent reporting the class of false bugs
generated by the unsound constraint solver while introducing minimal overhead to
the analysis. In particular, the empirical evaluation show that the bug
refutation extension can consistently reduce the analysis time if bugs are
removed, while the slow down when no bug is refuted is negligible.
We used five different solvers in our experiments (Z3, Boolector, MathSAT, Yices
and CVC4) and their performance is equivalent. Our refutation extension using Z3
is already part of the clang version 7 and the support for the other solvers
(Boolector, MathSAT, Yices and CVC4) is under review for clang version 8.

\section{Acknowledgements}

We would like to thank George Karpenkov and Artem Dergachev for the guidance
during the development of this project and, in particular, George Karpenkov for
the idea for the project.

%

\end{document}